\RequirePackage{ifpdf}

\documentclass{PoS}

\usepackage[authoryear,square]{natbib}
\bibpunct{(}{)}{;}{a}{}{,}
\usepackage{graphicx}
\usepackage{amsmath}
\usepackage{etoolbox}

\title{Cosmology with the SKA -- overview}

\ShortTitle{Cosmology with the SKA}

\author{\speaker{Roy Maartens}$^{1,2}$, Filipe B. Abdalla$^{3,4}$, Matt Jarvis$^{5,1}$, {Mario G. Santos}$^{1,6}$ -- ~~~~~~~~~~~~~~~~~~~~~~~~~~~~~~~~~~~~~
on behalf of the SKA Cosmology SWG\\
$^1$ Department of Physics, University of Western Cape, Cape Town 7535, South Africa\\
$^2$Institute of Cosmology \& Gravitation, University of Portsmouth,  Portsmouth PO1 3FX, UK
\\ $^3$Department of Physics \& Astronomy, University College London, London WC1E 6BT, UK\\
$^4$Department of Physics \& Electronics, Rhodes University, Grahamstown 6140, South Africa\\
$^5$Astrophysics, University of Oxford, Oxford OX1 3RH, UK \\
$^6$SKA South Africa, The Park, Park Road, Cape Town 7405, South Africa\\

E-mail: \email{roy.maartens@gmail.com}}

\abstract{The new frontier of cosmology will be led by three-dimensional surveys of the large-scale structure of the Universe. Based on its all-sky surveys and redshift depth, the SKA is destined to revolutionize cosmology, in combination with future optical/ infrared surveys such as Euclid and LSST. Furthermore, we will not have to wait for the full deployment of the SKA in order to see transformational science. In the first phase of deployment (SKA1), all-sky HI intensity mapping surveys and all-sky continuum surveys 
are forecast to be at the forefront on the major questions of cosmology. 
We give a broad overview of the major contributions predicted for the SKA. The SKA will not only deliver precision cosmology -- it will also probe the foundations of the standard model and open the door to new discoveries on large-scale features of the Universe. 
}

\FullConference{
Advancing Astrophysics with the Square Kilometre Array\\
June 8-13, 2014\\
Giardini Naxos, Italy}

\newcommand{\skipthis}[1]{}

\newcommand{\apjl}{ApJ}
\newcommand{\apjs}{ApJS}
\newcommand{\aap}{A \& A}

\newcommand{\be}{\begin{equation}}
\newcommand{\ee}{\end{equation}}
\newcommand{\bea}{\begin{eqnarray}}
\newcommand{\eea}{\end{eqnarray}}

\begin{document}


\section{SKA cosmology -- a new paradigm}

It has often been assumed in the past that the SKA would only be competitive in cosmology when the `billion galaxy survey' was completed -- i.e., when the full SKA (SKA2) was constructed. Only SKA2 can deliver the capacity for a spectroscopic survey of $\sim1$ billion HI galaxies.

Recently this view has been overturned. The billion galaxy survey will indeed be a game-changer in cosmology, in some senses the ultimate spectroscopic survey. However, well before this stage is reached, SKA1 will be able to deliver competitive and transformational cosmology. The main development that makes this possible is based on innovative ideas for deploying a new type of cosmological galaxy survey \citep{santos}: 
\begin{itemize}
\item
All-sky neutral hydrogen (HI) intensity mapping surveys that do not detect individual galaxies but only the integrated HI emission of galaxies in each pixel, together with very accurate redshifts at each tomographic slice. 
\end{itemize}
In addition, it has been recently realised that the radio continuum offers a novel probe of large-scale structure \citep{jarvis}:
\begin{itemize}
\item
All-sky radio continuum surveys that detect radio galaxies through their total emission out to very high redshift. 
\end{itemize}

\begin{figure*}
\begin{center}
  \includegraphics[width=0.9\textwidth]{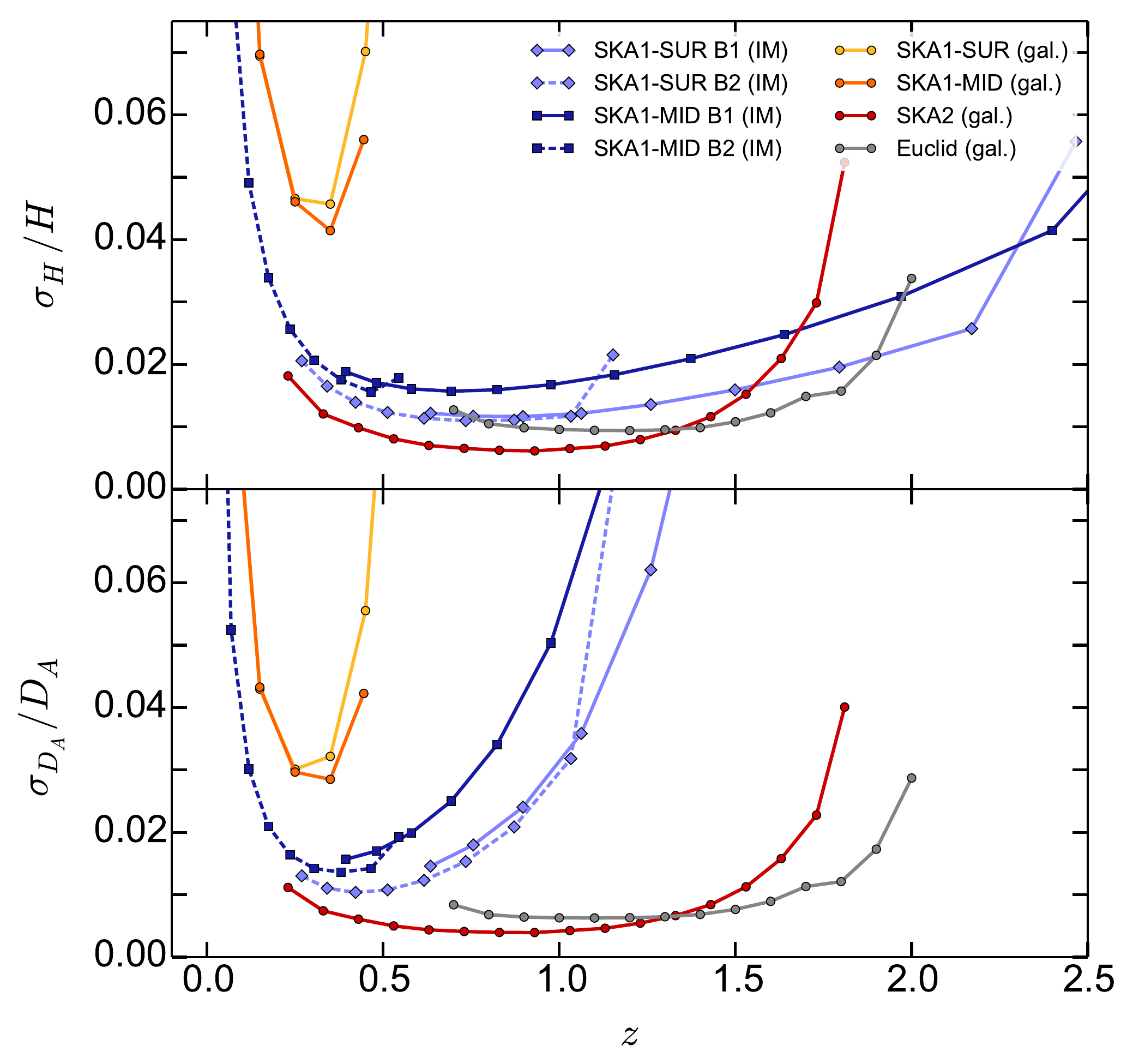}
   \caption{ Error on the Hubble rate and angular diameter distance from radial and transverse BAO measurements, showing performance of SKA HI surveys -- intensity mapping (IM) and  galaxy redshift surveys (gal). Euclid spectroscopic survey shown for comparison.   \citep{bull}} 
\label{gen}
\end{center}
\end{figure*}

Together with the `standard' HI galaxy redshift surveys, these radio surveys on  SKA1 and then SKA2 are destined to revolutionize cosmology. As an illustration of this, 
Fig. \ref{gen} shows forecasts for errors on measurements of the radial and transverse baryon acoustic oscillation (BAO) feature. In SKA1, the  HI intensity mapping surveys outperform the HI galaxy redshift survey -- and also current-generation optical galaxy surveys. 
The power of the HI galaxy survey in SKA2 is also evident. 

This chapter is a brief overview of the three chapters which review the science that the SKA can achive via three types of cosmological survey:
\begin{itemize} 
\item HI galaxy redshift surveys  \citep{abdalla} 
\item HI intensity mapping surveys  \citep{santos}
\item Radio continuum surveys \citep{jarvis}. 
\end{itemize}

These three review chapters highlight the main results in sixteen further chapters, which focus on specific science goals and techniques. The three review chapters also discuss the main technical challenges, which will not be covered here.

\section{The new frontier in cosmology}

Modern cosmology rests on the twin pillars of cosmic microwave background (CMB) surveys and surveys of the large-scale structure (LSS) of the Universe, supplemented by distance measures from type Ia supernovae (SNIa).  The CMB currently delivers the tightest constraints on the primordial Universe, since most of the information in CMB temperature and polarization anisotropies relates to the epoch of matter-radiation decoupling, $z\sim 1000$.  

There are CMB constraints on the low redshift Universe, principally via the lensing of the CMB by large-scale structure. The CMB can also contribute through cross-correlation with large-scale structure data, in the form of the integrated Sachs-Wolfe effect.
However, the main probe of the low-redshift Universe -- and especially of the critical question of the late-time acceleration of the Universe -- is the  large-scale distribution of matter (together with SNIa surveys). 
Galaxy surveys have not reached the levels of precision of the CMB. But major advances have been made, especially in measurements of the BAO scale.
\begin{figure*}
\begin{center}
  \includegraphics[width=0.8\textwidth]{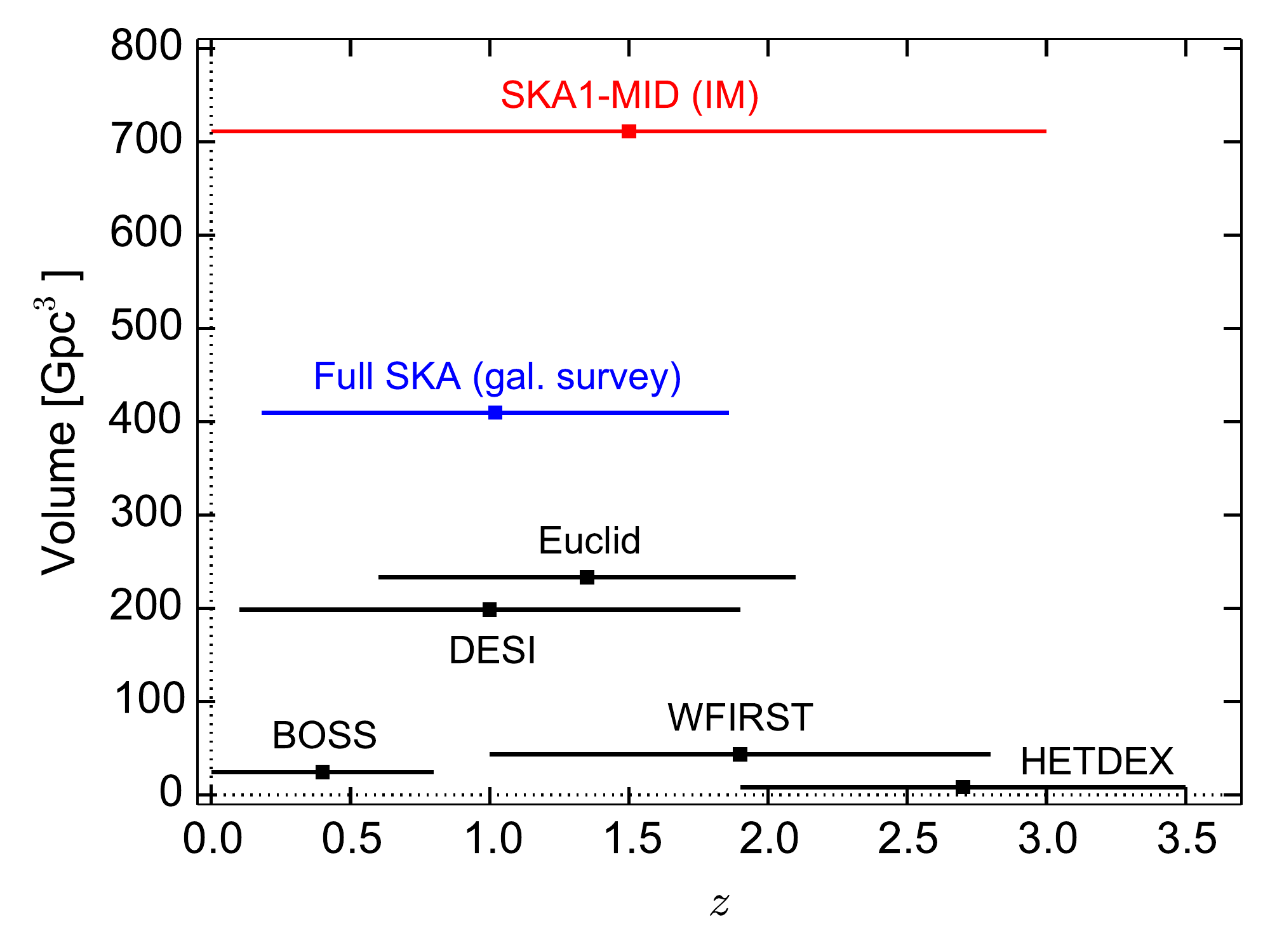}\vspace*{-0.5cm}
   \caption{ Survey volumes (at the midpoint of the redshift range) for various current and future surveys. \citep{santos}} 
\label{vol}
\end{center}
\end{figure*}

A new frontier of precision cosmology is emerging -- three-dimensional surveys of the LSS in the Universe. Current galaxy surveys do not yet cover both a wide area of sky and a significant redshift depth. This is what is needed for a high enough volume -- and thus a high enough number of modes -- for  next-generation precision cosmology. But future planned surveys, like Euclid, LSST and especially the SKA, will achieve both of these features. These LSS surveys will open up the new frontier of cosmology that can deliver precision at and beyond CMB levels. Indeed, the largest LSS surveys will be somewhat like performing the CMB survey over a range of redshifts.

The SKA will carry out higher volume surveys than ever before of the LSS of the Universe (see Fig. \ref{vol}). This will already be achieved in SKA1, with the HI intensity mapping surveys. With SKA2, the HI galaxy redshift survey will be the biggest ever spectroscopic galaxy survey.
In more detail:
\begin{itemize}
\item
HI intensity mapping surveys on SKA-MID and/ or SKA-SUR: $\sim$30,000 deg$^2$ out to $z\sim 3$ in SKA1. Even though individual galaxies are not detected, the resolution is more than adequate to measure the fluctuations needed for BAO and other large-scale features (such as primordial non-Gaussianity). Figures \ref{vol} and \ref{uls} show the enormous potential of these surveys for ultra-large-scale cosmology in SKA1.
\begin{figure*}
\begin{center}
  \includegraphics[width=0.8\textwidth]{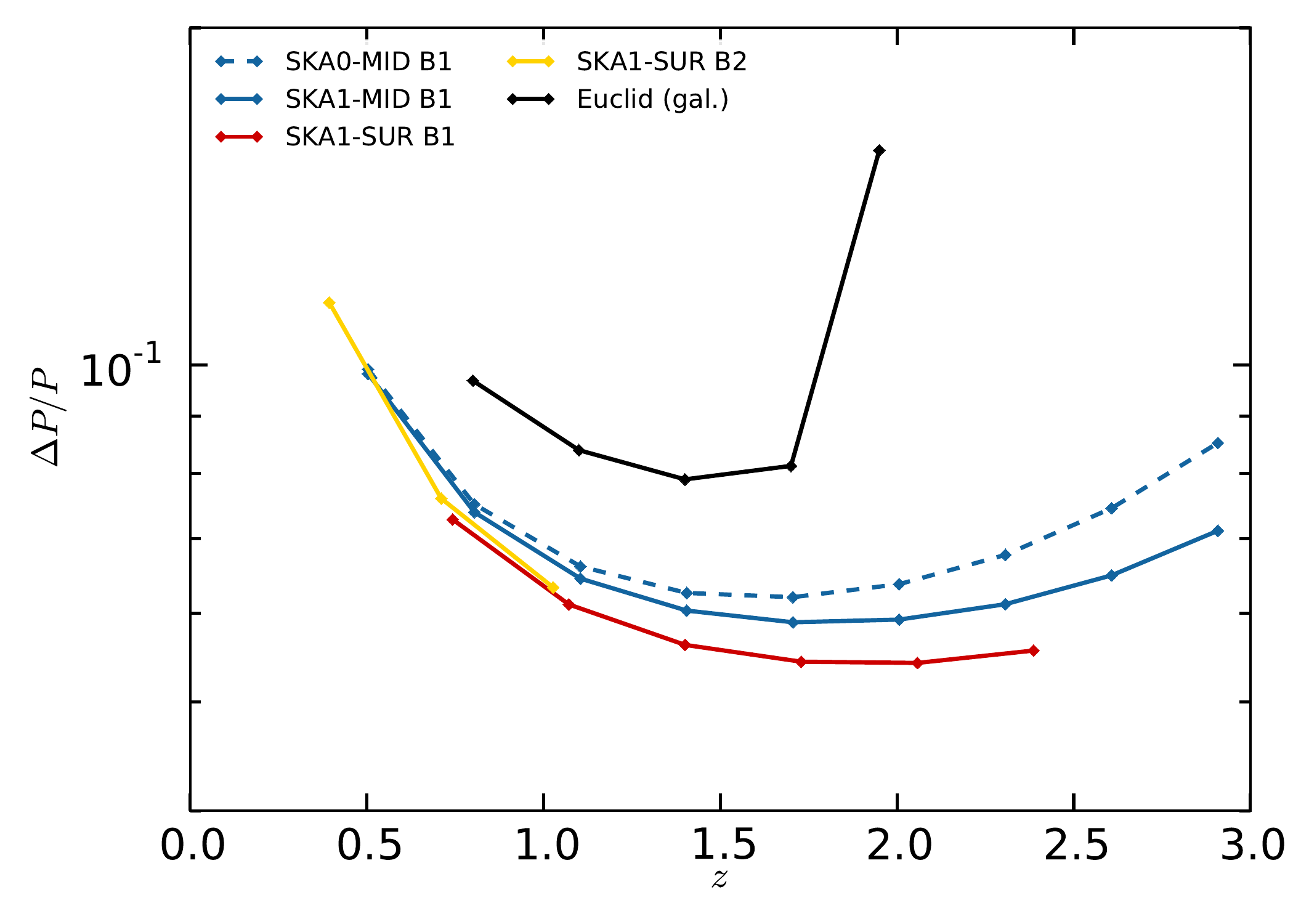}\vspace*{-0.5cm}
   \caption{ Errors on the power spectrum at very large scales, $k=0.01 h\,$Mpc$^{-1}$, achievable with SKA1 HI intensity mapping. SKA0-MID denotes early science on MID at $\sim50\%$ of the SKA1 specification.  Euclid spectroscopic survey is also shown for comparison. \citep{santos}} 
\label{uls}
\end{center}
\end{figure*}
\item
Radio continuum surveys on SKA-MID: $\sim$30,000 deg$^2$ out to $z\sim 6$, detecting $\sim 10^8$ galaxies in SKA1 and $\sim 10^9$ in SKA2. Although these are strictly two-dimensional surveys, they can be made partly three-dimensional by using redshift information from HI surveys or optical surveys. In addition, we can separate radio galaxy populations, leading to powerful applications of the multi-tracer technique.
\item
HI galaxy redshift surveys on SKA-MID and/ or SKA-SUR: $\sim$5,000 deg$^2$ out to $z\sim 0.7$, detecting $\sim 10^7$ galaxies (SKA1) and $\sim$30,000 deg$^2$ out to $z\sim 2$, detecting $\sim 10^9$ galaxies (SKA2 -- the billion galaxy survey). 
\item
Weak lensing surveys based on measuring the shapes of galaxies, starting with a $\sim$5,000 deg$^2$ survey on SKA1-MID and building up to a $\sim$30,000 deg$^2$ survey in SKA2. In SKA1, the lensing of the HI intensity mapping signal should also be detected (analogous to the detection by Planck of CMB lensing).
\end{itemize}

It is also possible to do some cosmology with SKA-LOW (see \citet{pritchard,santos}).

\section{Precision and discovery}

We often hear about the era of precision cosmology, in which observations are increasingly accurately tying down the handful of parameters that describe the standard `concordance' model, i.e. a spatially flat Friedmann model with cold dark matter (CDM) and dark energy in the form of vacuum energy (the cosmological constant $\Lambda$). The CMB and the evolution of large-scale structure are described by perturbations of the background model, and the origin of these perturbations is usually taken to be primordial inflation, driven by a simple single-field inflaton.

The concordance model has been extremely successful -- able to encompass a huge range of features and scales within in a single simple framework. It also has strong predictive power because of this simplicity. And so indeed a major part of cosmology is justifiably concerned with the precision determination of the parameters of the concordance model. The WMAP and Planck CMB surveys and the leading galaxy surveys, from 2dFGRS to WiggleZ, SDSS-III and DES, have laid the foundations for precision cosmology.
The SKA is designed to make a significant contribution to precision cosmology, based on the statistical power delivered by ultra-large volumes.

But there is more to cosmology than precision. Firstly, there are unresolved issues affecting the concordance model, the most important of which concerns the nature of dark energy. There is no satisfactory explanation for dark energy (in the form of the vacuum or otherwise) from fundamental physics, and we are reduced to testing various phenomenological models. Secondly, no model in physics is ever complete. No matter how good the model and the theory are, both will inevitably be replaced as new data and theoretical inconsistencies come to light. So we need to understand precision in a relative, and not absolute, sense. In fact, we have to go further -- physics involves also the testing of models with the aim of overturning them. 

The limits of precision cosmology are also the beginnings of `discovery cosmology'.  We cannot predict what new discoveries will emerge, but we need to orient towards the unexpected. We should not plan all of our observational tests in the framework of our current understanding of the concordance model, but rather allow for the possibility that new effects and new physics may be involved. One way to promote this is to devise and implement tests of fundamental features of the concordance model, in addition to the more routine tasks of measuring standard parameters. 

\section{Key science goals and SKA forecasts}

Cosmology requires LSS surveys that can accurately probe (a)~the expansion history and geometry of the Universe, and (b)~the growth of structure. The first are measured via the `standard ruler' imprinted in the correlation function by the BAO, with comoving scale $\sim 100 h^{-1}\,$Mpc. The growth of structure is described by the power spectrum of the observed density (or HI brightness temperature) contrast $\delta$, and observables derived from it, such as the growth rate $f=d\ln \delta/d\ln a$, which is measured via redshift space distortions (RSD). A weak lensing survey gives another independent probe of the LSS. 

The growth of structure provides a test for deviations from general relativity on large scales. Using the growth rate, we can measure $\gamma=\ln f/\ln \Omega_m$, looking for deviations from the general relativity value $\approx 0.55$. Using the density contrast and weak lensing measurements allows us in addition to test for `gravitational slip', i.e. a mismatch between the Newtonian and curvature perturbations that may signal deviations from general relativity.

The density contrast and weak lensing are mainly probes of the late-time accelerating Universe. LSS surveys can also probe the primordial Universe through the curvature and though large-scale correlations, which carry the signature of primordial non-Gaussianity, of general relativistic effects, and of possible deviations from statistical isotropy and homogeneity.

Some of the key questions at the forefront of cosmology today, where forecasts indicate that the SKA can be transformational, are briefly described below. Further details are given in the three review chapters  \citep{abdalla,jarvis,santos}.

\subsection{How were the primordial fluctuations generated?}
This fundamental question in cosmology requires inter-related tests via CMB and LSS surveys. 
The level of primordial non-Gaussianity is one of the most important discriminators of the primordial mechanism that generates cosmological fluctuations. The Planck CMB survey has achieved an error of $\sigma(f_{\rm NL}) =7.5$ (using the LSS convention), where $f_{\rm NL}$ is the primordial non-Gaussianity parameter. This is the current state of the art, and LSS surveys lag far behind. In the LSS power spectrum, the primordial non-Gaussian signal grows as $k^{-2}$, i.e. it is strongest on the largest scales. But these are also the scales where cosmic variance is an obstacle.

Ultra-large-volume surveys of the LSS are needed to surpass the CMB accuracy. Then we will be able to put pressure on the standard model (slow-roll single-field inflation), or give support to it. In the standard model,  
LSS surveys would see $f_{\rm NL}\simeq-2.2$, due to a nonlinear general relativistic correction \citep{Camera:AASKA14}. In order to implement this test, LSS surveys need to reach $\sigma(f_{\rm NL}) \lesssim 2$.

\begin{figure*}
\begin{center}
\includegraphics[width=7.5cm]{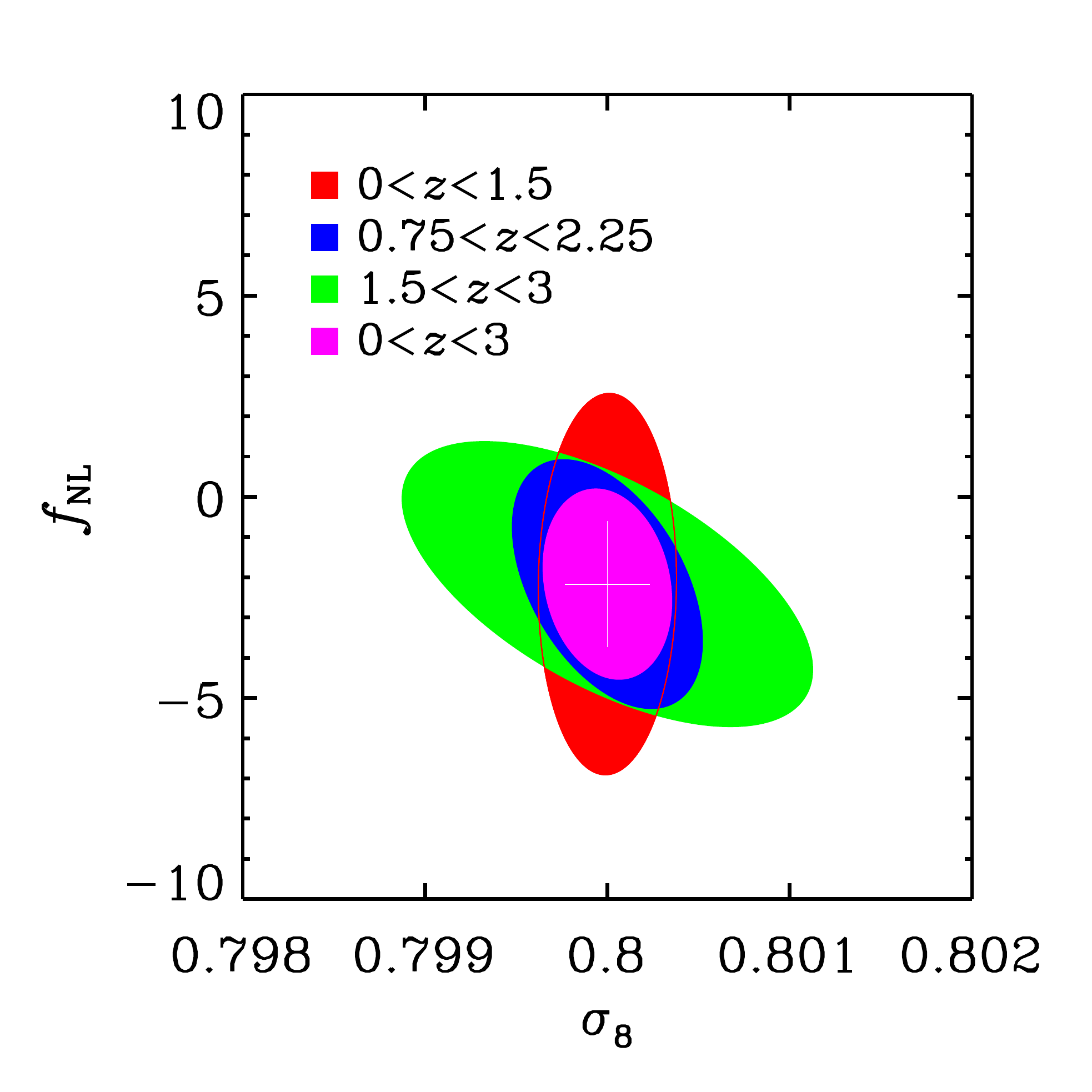} 
\includegraphics[width=7.5cm]{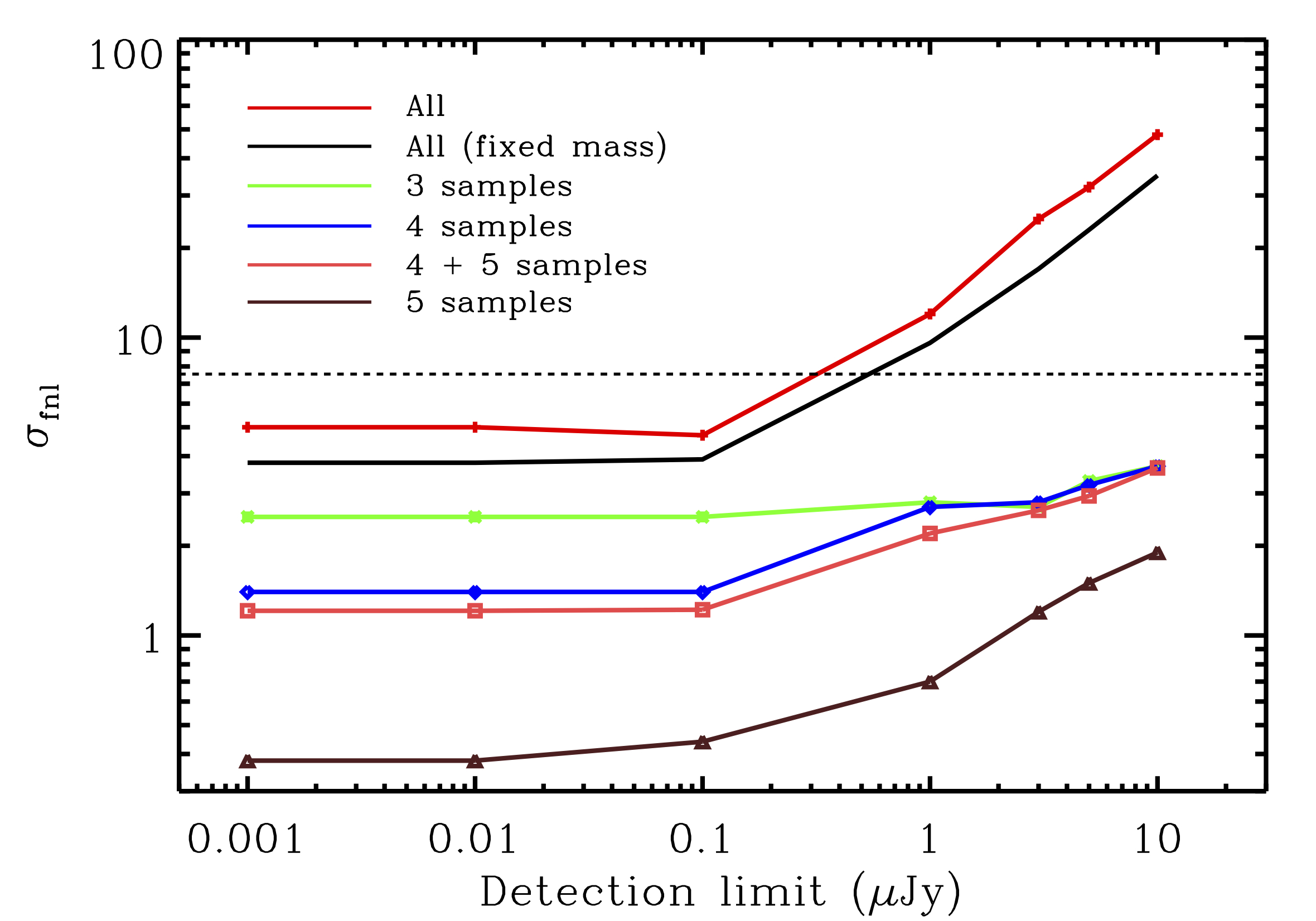}
\caption{Forecast errors on $f_{NL}$. {\em Left:} With the SKA2 galaxy redshift survey. \citep{Camera:AASKA14} {\em Right:} With continuum surveys at different sensitivities (including SKA1 and SKA2), using the multi-tracer method. 
 \citep{jarvis}
}\label{fig:NG}
\end{center}
\end{figure*}

In SKA Phase 1, intensity mapping surveys will achieve enormous volumes with accurate redshifts. The galaxy redshift surveys will not be competitive in SKA1 but SKA2 will deliver the best constraints for a single-tracer LSS survey. Finally, the continuum survey can also be effective if radio galaxy populations can be separated, allowing for the application of the powerful multi-tracer method that beats down cosmic variance on the large scales where non-Gaussianity is strongest. In summary, the forecasts are as follows:\\ For single-tracer measurements with HI surveys, we find \citep{santos,abdalla}
\bea
\sigma(f_{\rm NL}) = 2.3~\mbox{(SKA1 IM)},~~1.5~\mbox{(SKA2 GRS)}.
 \eea 
For multi-tracer measurements with continuum surveys \citep{jarvis}, 
\bea
\sigma(f_{\rm NL}) < 1~\mbox{(SKA2 Cont $+\,z$ from HI/optical)}.
 \eea 
See Fig. \ref{fig:NG} for the galaxy redshift and continuum surveys.

\subsection{What is driving the acceleration of the Universe?}
Perhaps the most fundamental question in cosmology is  -- what is driving the acceleration of the late-time Universe? Is it
\begin{itemize}
\item Vacuum energy ($\Lambda$, with $w=-1$) -- the simplest option (concordance model)?
\item Dynamical dark energy\footnote{Typically this is tested via the parametrization $w=w_0+w_a(1-a)$ or by using principal component analysis.}, with $w\neq -1$?
\item A weakening of gravity on very large scales, i.e. a breakdown in general relativity (`modified gravity')?
\end{itemize}

\begin{figure*}
\begin{center}
  \includegraphics[width=0.495\textwidth]{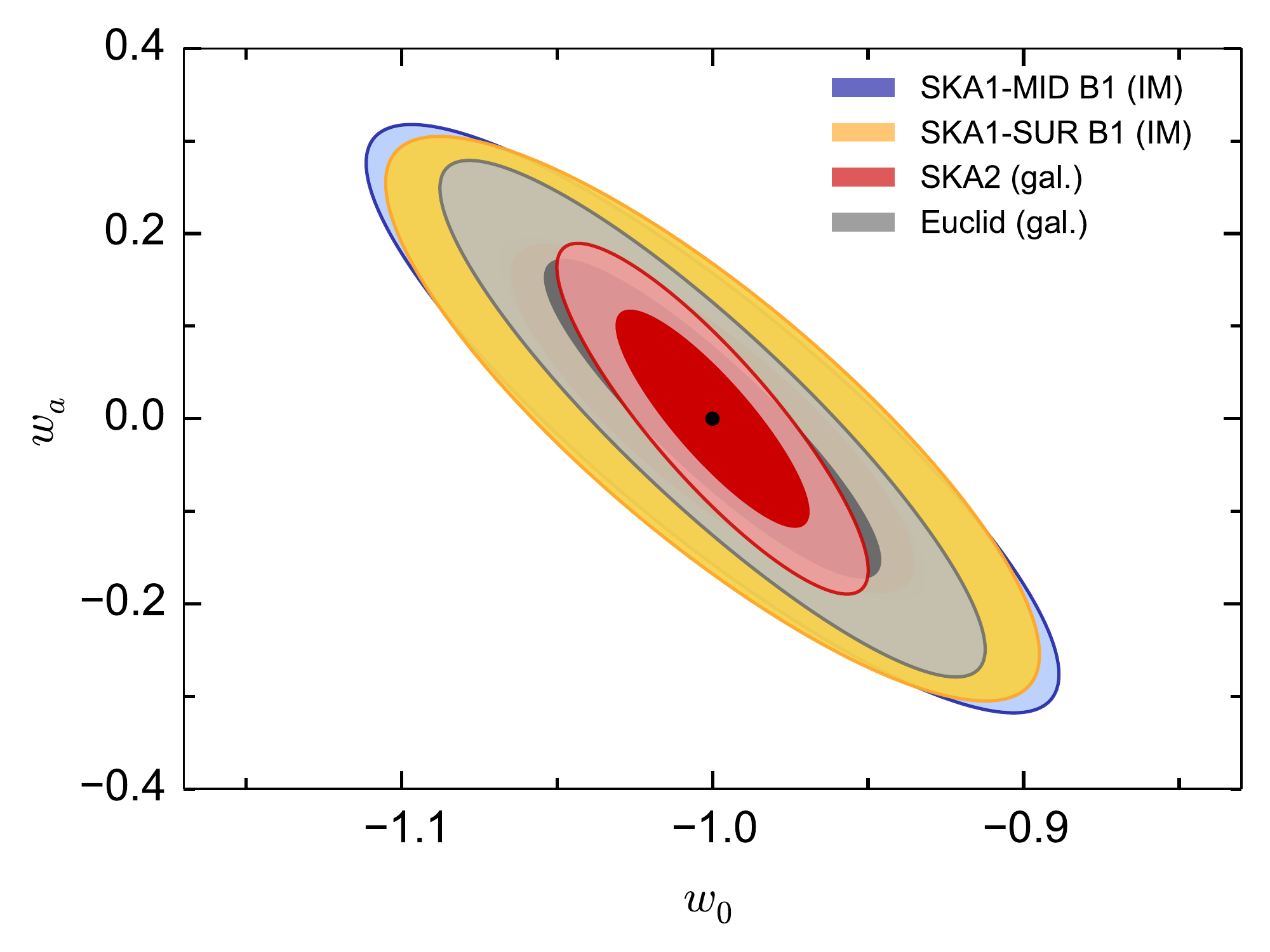}
  \includegraphics[width=0.495\textwidth]{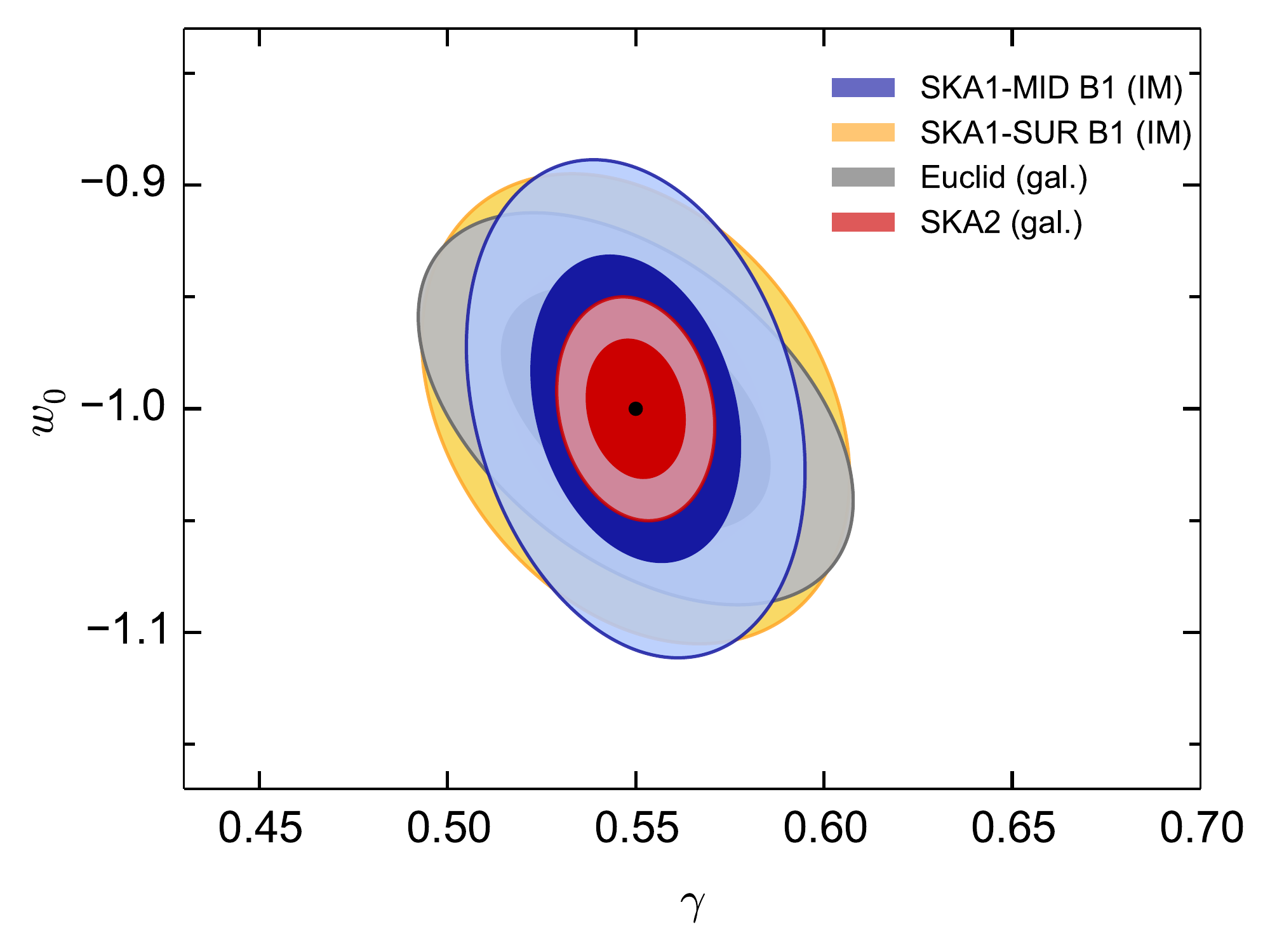}
  \caption{Constraints from RSD on the dark energy equation of state ({\em Left}) and on modified gravity (deviations from the GR growth rate) ({\em Right}), for SKA1 HI intensity mapping surveys, the SKA2 HI galaxy redshift survey and Euclid. (Planck and BOSS data inlcuded.) \citep{raccanelli}} 
\label{fig:dP}
\end{center}
\end{figure*}

Forecasts indicate that intensity mapping in SKA1 can outperform current-generation optical surveys, and is not far behind future optical spectroscopic surveys -- see Figs. \ref{gen} and \ref{fig:dP}. The latter figure also shows the game-changing power of SKA2 on the question of dark energy/ modified gravity. 

We are entering an era in which we can expect a definite answer to the acceleration question -- and the answer will have profound implications for our understanding of the Universe.
In order to succeed, we will need the combined power of the SKA and other LSS surveys like Euclid.  

\subsection{Is the Universe statistically isotropic and homogeneous?}
The concordance model, as well as dynamical dark energy and modified gravity models, are all based on the fundamental assumption that the Universe is statistically isotropic and homogeneous -- the `Cosmological Principle'. This principle should be interrogated by carefully designed observational tests. 

One critical feature of the standard assumption is that the dipole in the CMB -- accurately measured by Planck -- should match the dipole in the LSS, since both are predicted to originate from our motion relative to the common radiation/ matter frame in the background. Current contraints on the LSS dipole from the NVSS survey are too weak to answer the question of whether the dipoles agree. SKA all-sky continuum surveys will be able to constrain the LSS dipole at similar levels to Planck constraints on the CMB dipole. The error on the dipole direction $\theta$ is forecast to be \citep{schwarz}
\be
\sigma(\theta)\sim 5^\circ~~\mbox{(SKA1)},~~\sim1^\circ~~\mbox{(SKA2)}.
\ee
SKA2 is close to the Planck precision and improves on the NVSS constraints by a factor $\sim 100$. The outcome -- whether it is a confirmation or an overturning of the fundamental assumption -- will be a milestone for cosmology.

The angular two-point correlation function from SKA all-sky surveys can be used to probe the quadrupole and octupole of the LSS distribution. This will give a unique opportunity to test whether the anomalies in the low multipoles of the CMB are statistically significant or not. 

In addition to tests of statistical isotropy, we can test whether the matter distribution shows the same nearly scale-invariant behaviour, $n_s\sim 1$, on super-Hubble scales, as seen in the CMB. Various tests of statistical homogeneity have been devised, based on relationships between distances and Hubble rates as functions of redshift. BAO meaurements with the SKA (see Fig. \ref{gen}) will allow us to apply these tests at high enough accuracy to pose a real challenge to the standard model. 

\subsection{What is the distribution of matter on horizon scales and how does it evolve?}

Up to now, LSS surveys have not yet been able to probe the matter power spectrum  beyond the equality scale, $k\sim .01h\,$Mpc$^{-1}$. We have yet to confirm observationally the predicted turnaround in the power spectrum -- an important consistency test of the standard model. In addition, we need to probe further, and measure the power spectrum near the Hubble horizon. At higher redshifts, we can in principle probe super-horizon modes. On this basis we can confirm the predictions of the concordance model, or discover deviations. These deviations could be due to primordial non-Gaussianity or modified gravity -- or to some unexpected new feature. 

There are two basic requirements in order to succeed. Firstly, we need all-sky surveys with spectroscopy and deep redshift reach.  Secondly, we need the correct theoretical tools to analyse correlations on horizon scales and across large redshift distances -- i.e., we need to incorporate all the relativistic effects that correct the Newtonian-Kaiser approximation at high redshift and at horizon scales \citep{Camera:AASKA14}. 

SKA1 can deliver the all-sky spectroscopy via intensity mapping, up to $z\sim3$, while the galaxy redshift survey in SKA2 will reach $z\sim2$ with high angular resolution.
Already in SKA1, we will be able to
map for the first time the large-scale HI distribution in 3/4 of the universe,  from today all the way back to redshifts where the concordance model predicts no effect of dark energy. SKA-LOW will allow us to probe even larger scales in the Epoch of Reionization, as  the angular scale of the horizon becomes smaller and smaller at very high redshift.

We will then be able to test the concordance predictions at high $z$. Furthermore, we will be able for the first time, to extend the tests of general relativity to horizon scales. 

\subsection{What is the curvature of the Universe?}
The concordance model has zero spatial curvature, $\Omega_K=0$, but small nonzero curvature is allowed by current data. LSS surveys with huge volume are needed to move beyond CMB accuracy ($|\Omega_K|\lesssim 10^{-2}$ from Planck) and determine whether the curvature is above the perturbative level of $O(10^{-5})$. A detection of non-perturbative curvature could rule out many inflation models. SKA1 intensity mapping surveys have huge volume and accurate spectroscopy, and they reach well beyond the redshifts that are affected by acceleration, thus helping to break degeneracies arising from dark energy. This leads to a predicted accuracy of $|\Omega_K|< 10^{-3}$, using BAO measurements.

\subsection{A revolution in weak lensing}

The measurement of galaxy shapes provides a statistical estimate of the weak gravitational lensing shear due to the intervening LSS, and this gives a powerful probe of the total matter and its distribution. Cosmological weak lensing surveys have so far been in the optical, but in principle they can also be carried out in the radio. The SKA offers the possibility of 
the first ever weak lensing survey in the radio that can deliver cosmological precision.  Such surveys would map the dark matter and dark energy/ modified gravity in an entirely new and independent way from their optical counterparts. A cross-correlation of radio and optical would offer a major reduction in sytematics that could significantly enhance the precision of weak lensing.

\begin{figure*}
\includegraphics[width=15cm]{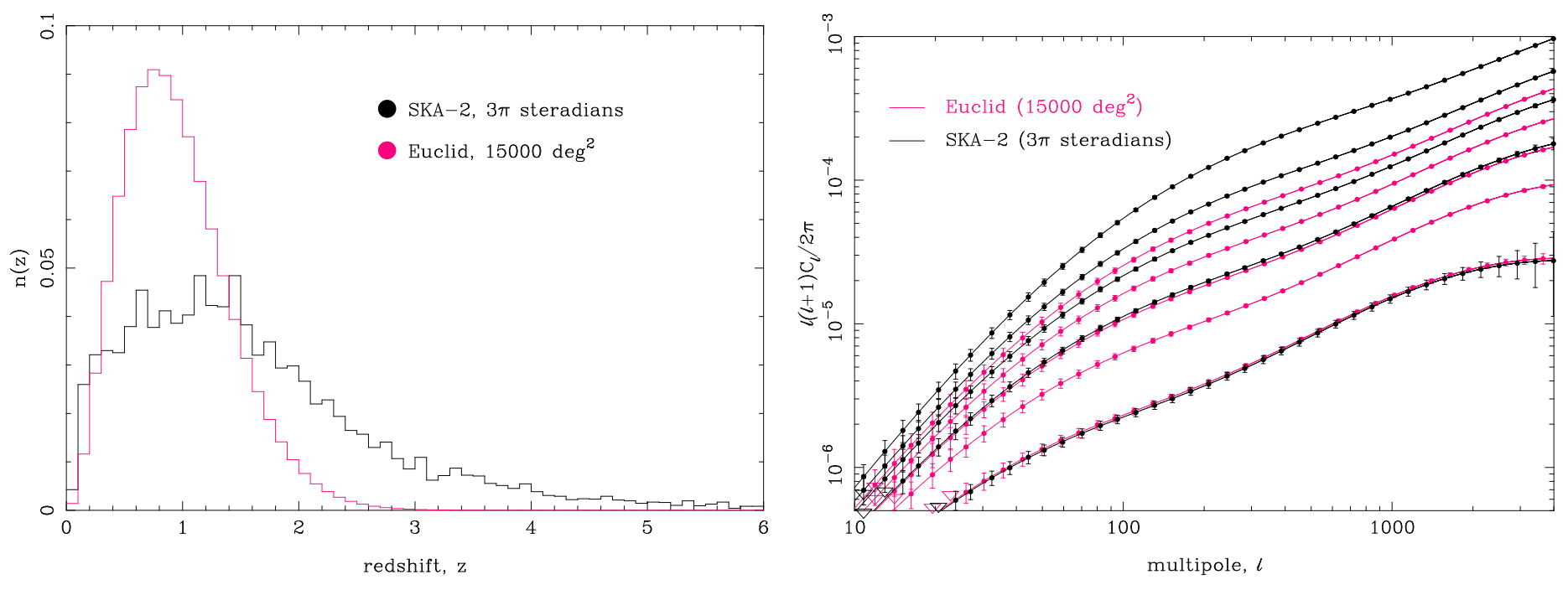}
\caption{{\em Left:} Redshift distribution of sources for a 30,000 deg$^2$ weak lensing survey with SKA2, and for the 15,000~deg$^2$ survey with Euclid. {\em Right:} The corresponding constraints on a 5-bin tomographic auto-power spectrum analysis. \citep{jarvis}}\label{fig:lensing}
\end{figure*}

A continuum survey in SKA1, covering 5,000 deg$^2$, would provide the foundation to prepare for a 30,000 deg$^2$ survey with SKA2, which is predicted to have transformational potential 
in combination with Euclid -- as shown in Fig. \ref{fig:lensing}.

\section{Conclusions}

We have given a brief overview of the major science goals and capabilities of the SKA in cosmology. Further details on these science topics are to be found in the three review chapters, covering the three types of survey \citep{abdalla,jarvis,santos}. Among the novel science goals described in those reviews and not covered here, are: 
\begin{itemize}
\item Using the topology of the HI distribution as a cosmological probe (SKA1 and SKA2);
\item Constraining primordial non-Gaussianity through the HI bispectrum (SKA1 and SKA2); 
\item Measuring the redshift drift of sources -- i.e. the real-time tracking of the change in redshift of a source -- as a probe of acceleration (SKA2).
\end{itemize}

The review chapters also examine in detail the crucial technical issues which were not discussed here, such as foreground contamination and instrumental systematics.

SKA is a facility that is planned to have a $\sim50$ year lifetime. It will be the premier facility for spectroscopy up to $z\sim3$,  and will not be surpassed by other spectroscopic surveys in volume, since HI is common up to high redshifts. On longer time-scales, one can envisage significant improvements to the forecasts described here. However, we have focused on the shorter term. In particular, we have shown that the SKA can deliver not only competitive, but transformational, science even in Phase 1 of the deployment, years before full deployment. With full deployment comes the ultimate `billion galaxy survey'.

In this overview, we have focused on the potential of SKA itself, and have not discussed the additional power arising from combining the SKA with Euclid, LSST and other future optical/ infrared surveys. Indeed, this combination will allow us to beat down experimental systematics and cosmic variance, delivering greater power than each survey on its own.

The science goals that we discussed included topics falling under the heading of `precision cosmology', i.e. tying down with ever greater accuracy the key parameters of the concordance model. But in addition, there are other goals that probe the foundations of the concordance model, looking for deviations -- both those theoretically foreseen and those that will be unexpected discoveries.


\bibliographystyle{apj}

\bibliography{overv}

\end{document}